\documentclass[9pt,letterpaper,technote]{IEEEtran}

\ifCLASSOPTIONcompsoc
  \usepackage[nocompress]{cite}
\else
  \usepackage{cite}
\fi

\usepackage{makeidx} 
\usepackage{subfigure}%
\usepackage{graphicx,tipa}
\usepackage[table]{xcolor}
\usepackage{booktabs,threeparttable}
\usepackage{epstopdf}
\usepackage{amsmath,amsthm}
\usepackage[T1]{fontenc}
\usepackage[utf8]{inputenc}
\usepackage{cuted} 
\usepackage{hyperref}
\usepackage{enumitem} 

\usepackage{float}
\floatstyle{plain}
\restylefloat{table}
\usepackage{multirow}

\newcommand{\ignore}[1]{}

\newcommand{\arxivOnly}[1]{#1}
\newcommand{\TKDEOnly}[1]{}

\DeclareMathOperator*{\argmax}{argmax}

\renewcommand{\baselinestretch}{0.96}

\begin{document}

\TKDEOnly{
\title{Comparing Alternative Route Planning 
Techniques: A Comparative User Study on Melbourne, Dhaka and Copenhagen Road Networks}
}

\arxivOnly{
\title{Comparing Alternative Route Planning 
Techniques: A Comparative User Study on Melbourne, Dhaka and Copenhagen Road Networks}
}

\author{Lingxiao~Li,
        Muhammad~Aamir~Cheema,~\IEEEmembership{Senior Member, IEEE},
        Hua~Lu,~\IEEEmembership{Senior Member, IEEE},
        Mohammed~Eunus~Ali,
        Adel~N.~Toosi
\thanks{L. Li, M. A. Cheema and A. N. Toosi are with the Faculty of Information Technology, Monash University, Australia.\protect\\
E-mails: \{lingxiao.li, aamir.cheema, adel.n.toosi\}@monash.edu\protect\\}%

\thanks{H. Lu is with the Department of People and Technology, Roskilde University, Denmark. \protect\\
E-mail: luhua@ruc.dk\protect\\}%

\thanks{M. E. Ali is with the Department of Computer Science and Engineering, Bangladesh University of Engineering and Technology, Bangladesh.\protect\\
E-mail: eunus@cse.buet.ac.bd }
}


%



\IEEEtitleabstractindextext{%
\begin{abstract}
Many modern navigation systems and map-based services do not only provide the fastest route from a source location $s$ to a target location $t$ but also provide a few alternative routes to the users as more options to choose from. Consequently, computing alternative paths has received significant research attention. However, it is unclear which of the existing approaches generates alternative routes of better quality because the quality of these alternatives is mostly subjective. Motivated by this, in this paper, we present a user study conducted on the road networks of Melbourne, Dhaka and Copenhagen that compares the quality (as perceived by the users) of the alternative routes generated by four of the most popular existing approaches including the routes provided by Google Maps. We also present a web-based demo system that can be accessed using any internet enabled device and allows users to see the alternative routes generated by the four approaches for any pair of selected source and target. We report the average ratings received by the four approaches and our statistical analysis shows that there  is  no  credible  evidence  that  the  four  approaches  receive different ratings on average.  We also discuss the limitations of this user study and recommend the readers to interpret these results with caution
because certain factors may have affected the participants' ratings.
\end{abstract}

}

\maketitle

\IEEEdisplaynontitleabstractindextext

%
\IEEEpeerreviewmaketitle

\section{Introduction}

Given a source location $s$ and a target location $t$ in a graph, a shortest path query~\cite{shao2016vip,ouyang2020efficient,abraham2011hub,shao2020efficiently} returns the path from $s$ to $t$ with the minimum total weight (e.g., travel time, distance). The shortest path query is one of the most fundamental queries in graphs and has applications in a wide variety of
domains. Throughout this paper, we use path and route interchangeably. The shortest path query has been very well studied in road networks (e.g., see~\cite{li2017experimental,wu2012shortest}) where a user may issue the shortest path query to find a route to travel
from one location to the other. Since the shortest path may not always match a user's traveling choices, modern map-based systems often provide several alternative routes so that
the user can choose a route that they find most suitable. It is critical for the
effectiveness of the alternative routes that these routes are significantly different from each other and are meaningful (e.g.,
without unnecessary  detours). 

Inspired by the importance of alternative routes, in the past several years, a large body of research has focused on computing alternative routes~\cite{jones2012method,abraham2013alternative,chen2007reliable, akgun2000finding,chondrogiannis2020finding,chondrogiannis2018finding,liu2017finding}.
Intuitively, the alternative routes reported to the users must be meaningful/natural and significantly different from each other. Different existing techniques propose different intuitive ideas to define and generate good quality alternative routes. 
However,  there is no agreed definition of what constitutes a set of ``good'' alternative routes~\cite{Li2020Alternative}. This is because the ``goodness'' of the alternative routes is mostly subjective and it is not trivial to define quantitative measures to evaluate the quality of routes.
Since the existing techniques define and generate alternative routes using inherently different approaches, it is not clear which of these techniques generates alternative routes which are perceived by the users to be of better quality.
 Surprisingly, despite a large body of research on generating alternative routes, there does not exist any systematic study that evaluates the existing techniques on the quality of the routes generated by these techniques.
 This makes it hard for the researchers, open-source community and developers to decide which techniques to use (or extend upon) for their own research, open-source systems and products etc.  Therefore, it is critical to conduct a user study to compare the perceived quality of the alternative routes generated by the existing techniques.

To fill this gap, in this paper, we present the first user study that compares four popular techniques including Google Maps which is among the most widely used commercial solutions providing alternative routes. Specifically, we create a web-based demo system that asks users to select source and target locations on the road networks of Melbourne, Dhaka or  Copenhagen, three demographically diverse cities having widely different population, traffic congestion, and density etc. It then displays up to $3$ routes generated by each of the following four techniques: Google Maps, Plateaus~\cite{jones2012method,abraham2013alternative}, Penalty~\cite{chen2007reliable, akgun2000finding} and Dissimilarity~\cite{chondrogiannis2020finding,chondrogiannis2018finding,liu2017finding}. The users are then asked to provide a rating from 1-5 (higher the better) for each of the four approaches. 
In total, we received $520$ responses ($237$ for Melbourne, $155$ for Dhaka and $128$ for Copenhagen).  We show mean rating and standard deviation for each of the four approaches for different groups of respondents (residents and non-residents). Also, we show the mean rating based on the lengths of the routes (small routes, medium routes and long routes). 
A one-way ANOVA test shows that the results are not statistically significant, i.e., there is no evidence that the approaches receive different ratings on average. We also evaluate the approaches on the similarity of their reported routes and we find that the routes reported by Dissimilarity are least similar to each other which is mainly because it specifically prunes the routes that have similarity above a certain threshold.  

We remark that the data used by Google Maps\footnote{Google Maps uses real-time and/or historical traffic data to compute the routes. This data is not made publicly available. Therefore, we were unable to use Google Maps data for all four approaches. Also, it is not possible to enforce Google Maps to generate alternative routes using the OpenStreetMap data used by the other approaches.} to compute the alternative routes is different from the OpenStreetMap (OSM) data used by the other three approaches. We provide the details of how this may have impacted the participants' ratings. We also list some other limitations of this user study that are beyond our control.  Nevertheless, we believe that it is fair to conclude that the three publicly available techniques (Plateaus, Penalty and Dissimilarity) produce alternative routes of quality comparable to the routes generated by Google Maps and can be used in commercial products and open-source systems to report high quality alternative routes.

Below we summarize the contributions we make in this paper.

\begin{itemize}
    \item To the best of our knowledge, we are the first to conduct a systematic user study comparing the path quality of some of the most popular approaches to generate alternative routes. 
    
    \item Our web-based demonstration system can be used by anyone with internet access to visualise the alternative routes and evaluate these approaches. Furthermore, we make our source code publicly available\footnote{\url{https://bitbucket.org/lingxiao29/demoapp/}} to facilitate extension to the demo system and to conduct further studies.  
    
    \item For fairness, we also list certain limitations of this study that were beyond our control and may have impacted the ratings received by each approach. Nevertheless, the results of this user study suggest that the existing published techniques perform reasonably well compared to Google Maps which is among the most widely used commercial approach.
\end{itemize}

The rest of the paper is organized as follows. In Section~\ref{sec:related}, we describe some of the most popular existing techniques to compute alternative paths including the three approaches we use in the user study. The details of our web-based demo system are presented in Section~\ref{sec:system}. The details of our user study and its  results are presented in Section~\ref{sec:experiments}. Some limitations of the user study are also discussed in Section~\ref{sec:experiments}. Section~\ref{sec:conclusion} concludes this paper.

\section{Related Work}\label{sec:related}

Computing alternative routes has received significant research in the past decade or so~\cite{jones2012method,abraham2013alternative,chen2007reliable, akgun2000finding,chondrogiannis2020finding,chondrogiannis2018finding,liu2017finding}. In this section, we briefly describe the existing techniques to generate alternative routes. However, we focus on presenting the details of the three most popular existing approaches widely used in literature that we compare in this user study: Penalty~\cite{chen2007reliable, akgun2000finding,bader2011alternative,DBLP:conf/atmos/KobitzschRS13,DBLP:conf/gis/ChengGZPW19}, Plateaus~\cite{jones2012method,abraham2013alternative,kobitzsch2013alternative,DBLP:conf/atmos/ParaskevopoulosZ13,dobler2016computation,Li2020Alternative} and Dissimilarity~\cite{chondrogiannis2020finding,chondrogiannis2018finding,chondrogiannis2015alternative,liu2017finding}. 

\subsection{Penalty}

The basic idea behind this approach~\cite{akgun2000finding,chen2007reliable,DBLP:conf/gis/ChengGZPW19} is to iteratively compute shortest paths and, after each iteration, apply a penalty on each edge of the shortest path found in the previous iteration (by increasing its weight by a certain factor). Since the edge weights of the previous shortest paths are increased, it is likely that the new shortest path found on the graph will be different from the previous path(s). The algorithm stops when $k$ shortest paths are retrieved. 

This approach does not guarantee that the paths are ``signficiantly'' different from each other or are meaningful (e.g., without small detours). However, we observe that, in practice, the routes generated by this approach turn out to be pretty good in most of the cases. This is mainly because, on typical road networks, there exists several different paths from $s$ to $t$ with very similar traveling time. Thus, when the algorithm applies penalty to one of these paths, it naturally tends to select the other significantly different paths.  Furthermore, after retrieving each path, the algorithm can specifically apply  additional filtering criteria to remove the paths that fail to meet certain requirements, e.g., the paths that are too similar to existing paths or have  detours can be ignored.

\subsection{Plateaus}

\begin{figure}[t]
	\centering
  \subfigure[$T_f$ rooted at Cambridge]{\includegraphics[scale=0.275]{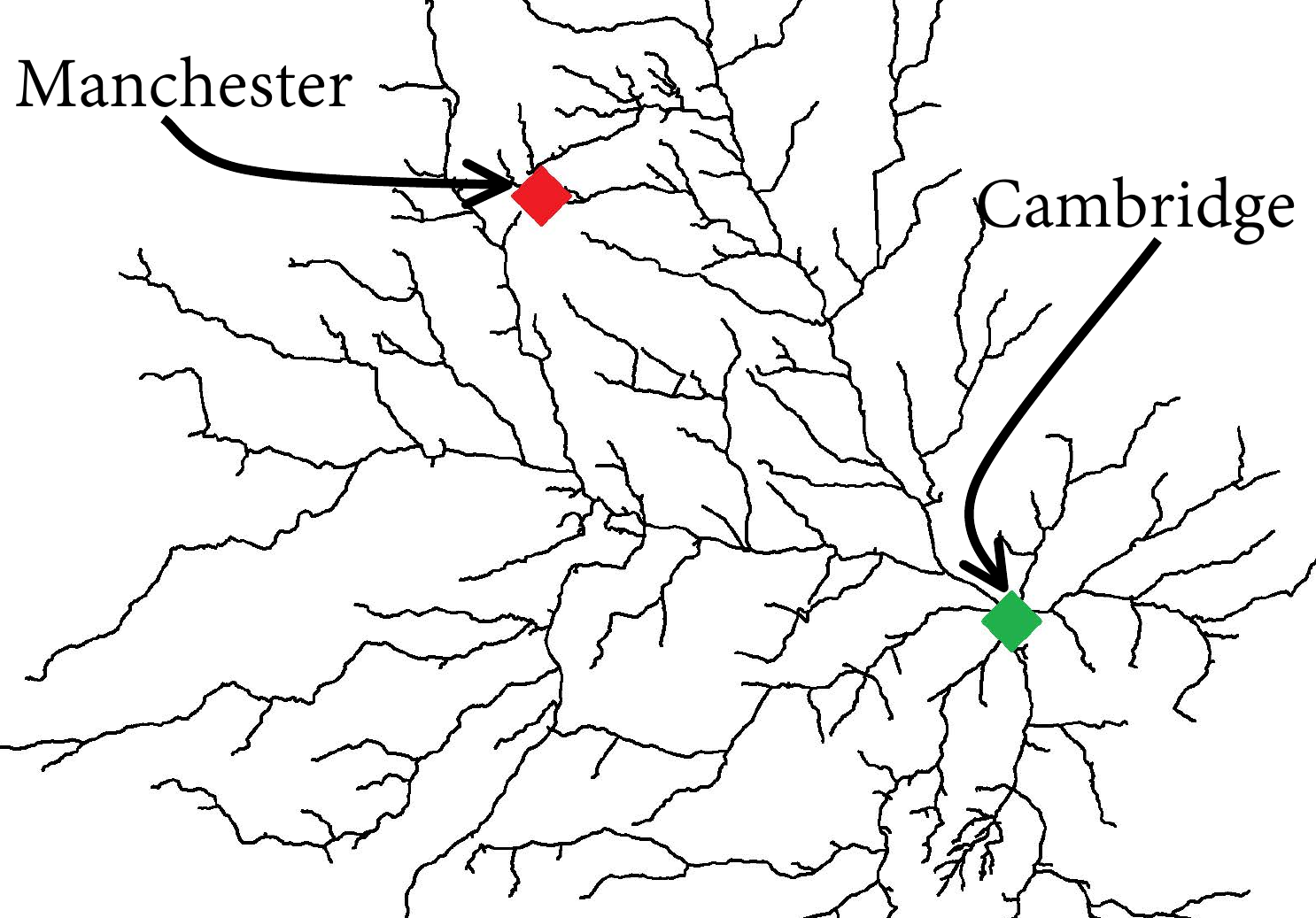}}
  \subfigure[$T_b$ rooted at Manchester]{\includegraphics[scale=0.275]{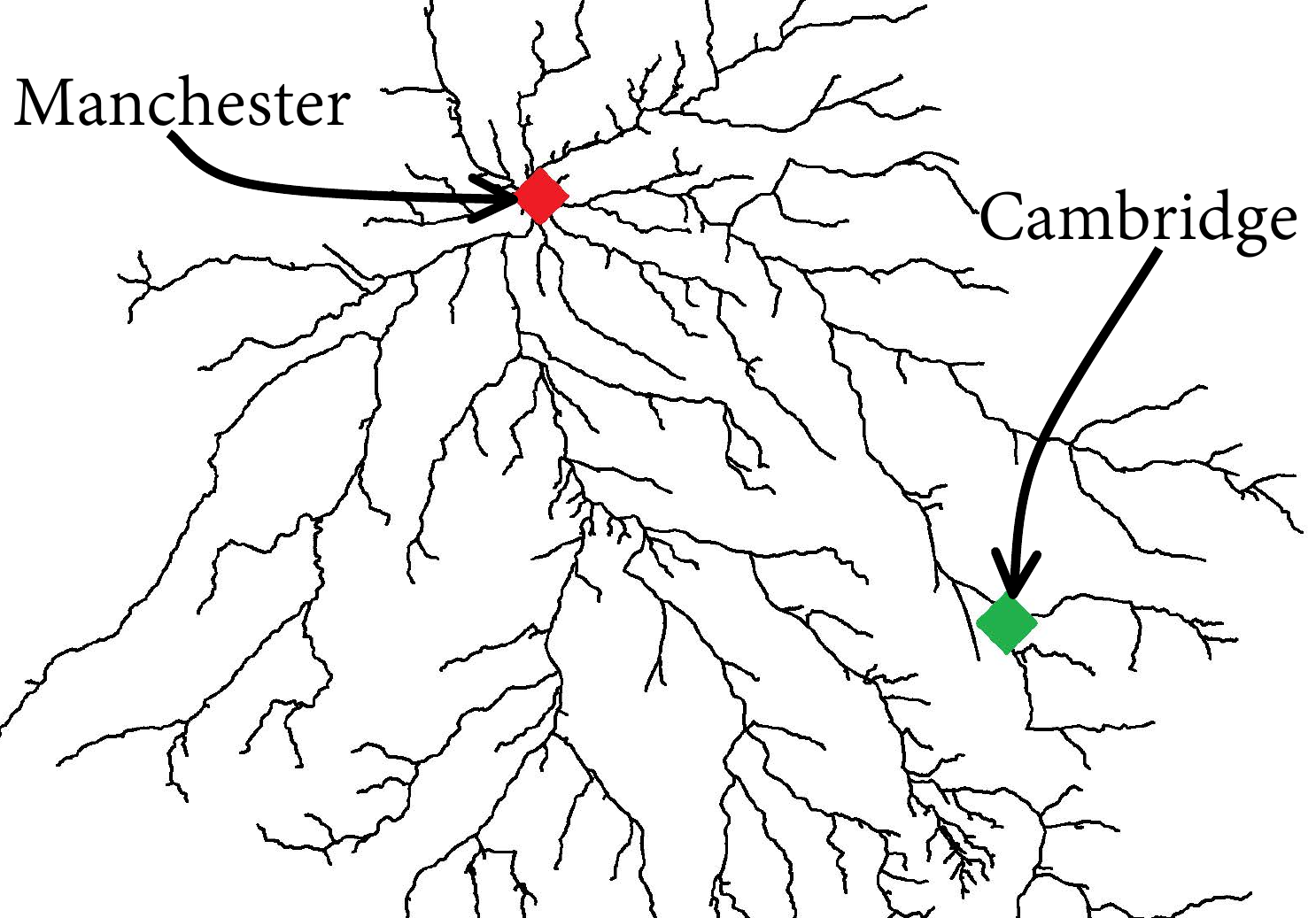}}
  \subfigure[Tree Join]{\includegraphics[scale=0.275]{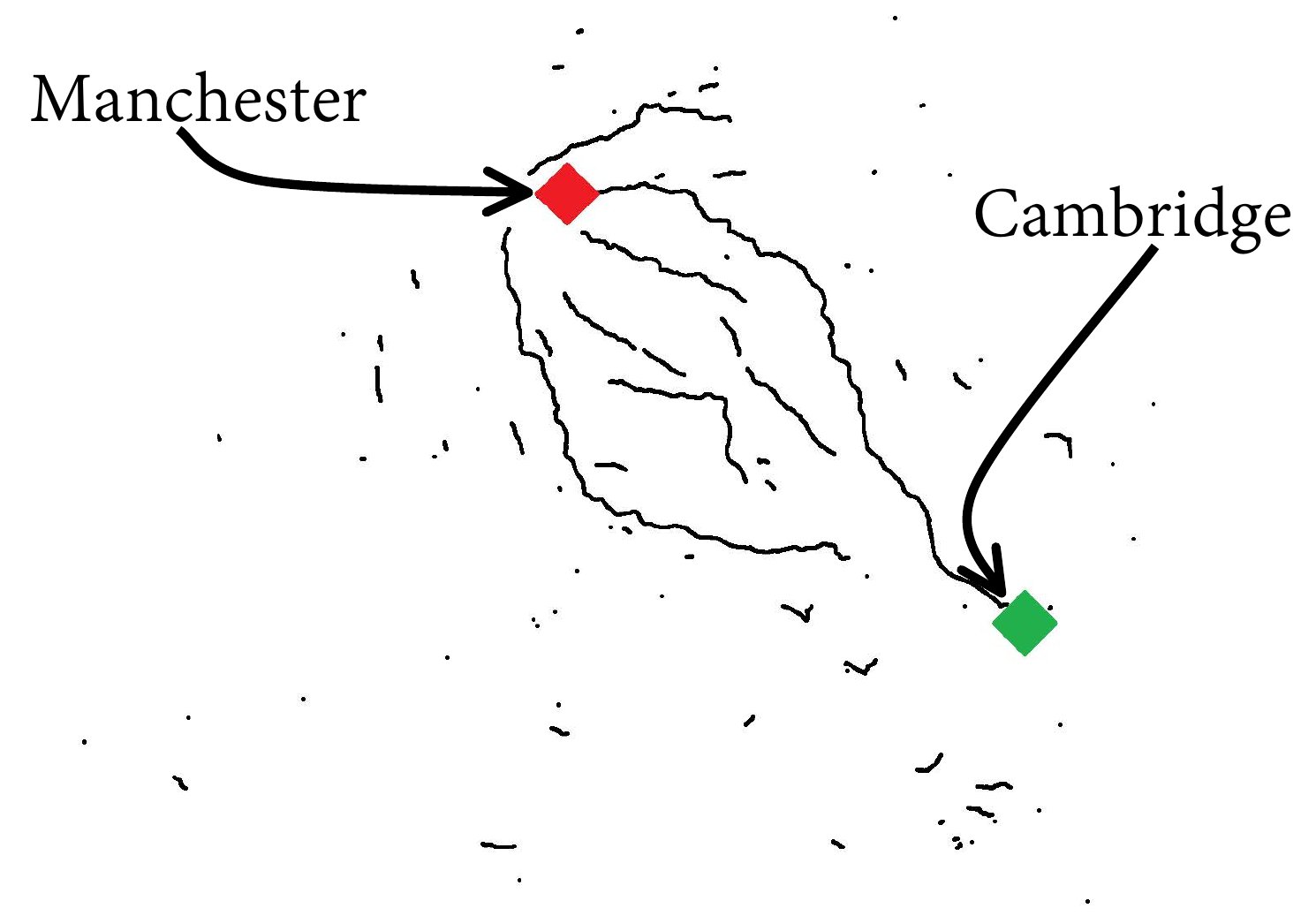}}
  \subfigure[Plateaus-based paths]{\includegraphics[scale=0.275]{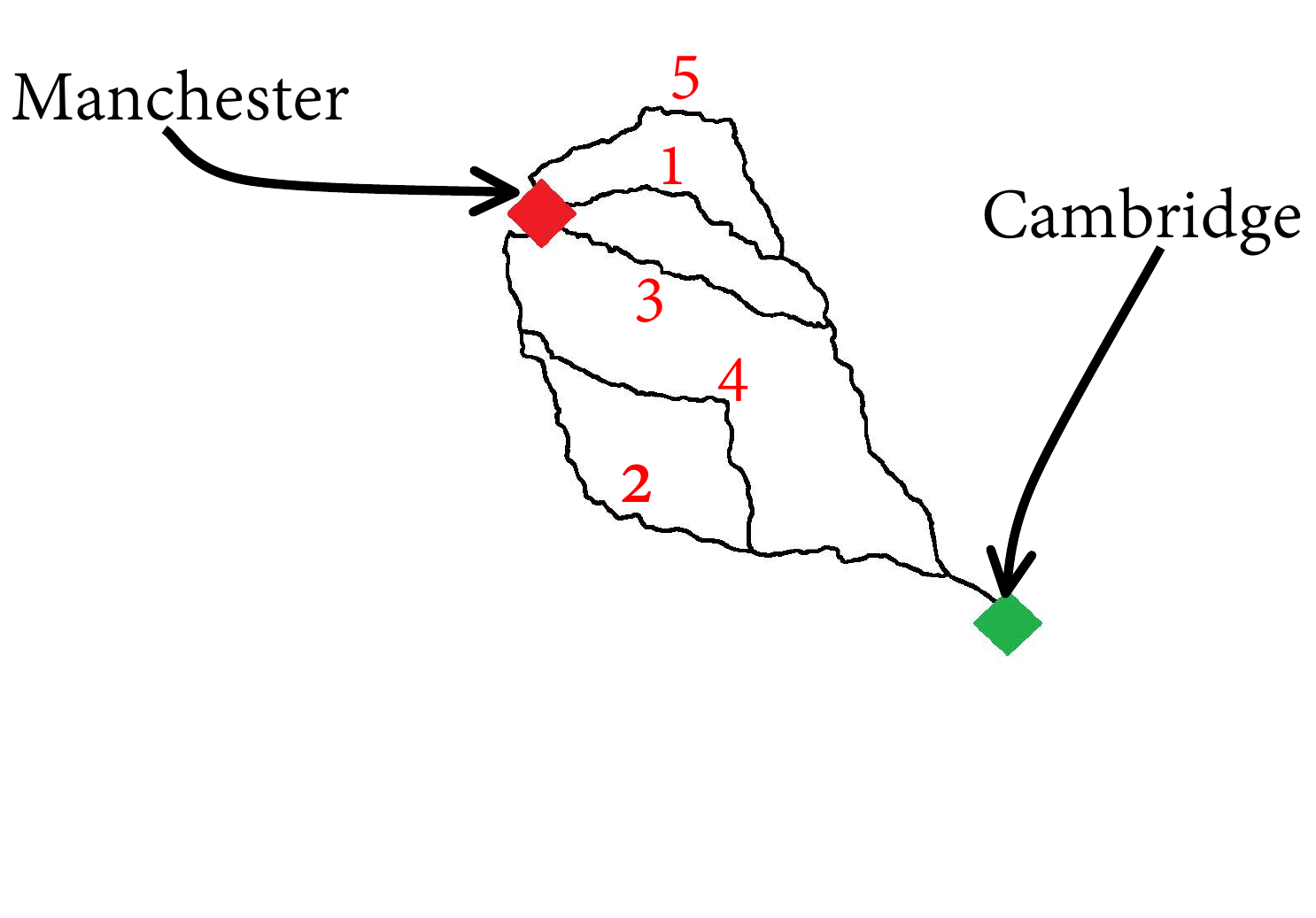}}
  \vspace{-3mm}
  \caption{Illustration of generating alternative paths using plateaus. Images adapted from \protect\cite{jones2012method} with permission.}
  \label{fig:pl_exp}
\end{figure}

The technique to generate alternative paths using plateaus  was developed~\cite{jones2012method} by Cotares Limited for their routing engine Choice Routing.  
 We use Fig.~\ref{fig:pl_exp} to illustrate how alternative paths from Cambridge (source $s$) to Manchester (target $t$) are generated using plateaus. First, two shortest path trees are generated: a forward shortest path tree $T_f$ rooted at $s$ (see Fig.~\ref{fig:pl_exp}(a)); and a backward shortest path tree $T_b$ rooted at $t$ (Fig.~\ref{fig:pl_exp}(b)). Then, the two trees $T_f$ and $T_b$ are joined to obtain the branches common in both trees. These common branches are called plateaus. Fig.~\ref{fig:pl_exp}(c) shows some of the most prominent plateaus.
 It was noted~\cite{jones2012method} that longer plateaus result in more meaningful alternative paths.
  Therefore, top-$k$ plateaus are selected based on their lengths. Let $u$ and $v$ be two ends of a plateau $pl(u,v)$ where $u$ is the end closer to the source and $v$ is the end closer to the target. Each plateau $pl(u,v)$ is used to generate an alternative path by appending the shortest paths from $s$ to $u$ and $v$ to $t$ to the plateau. Fig.~\ref{fig:pl_exp}(d) shows five alternative paths generated using the five longest plateaus from Fig.~\ref{fig:pl_exp}(c).

It was shown that the alternative paths generated using plateaus are \emph{local optimal}~\cite{abraham2013alternative}. Furthermore, the plateaus do not intersect each other. Thus, two paths generated using longer plateaus are expected to have a smaller overlap (i.e., lower similarity). Thus, the paths generated using the $k$ longest plateaus are likely to be more dissimilar to each other (although this cannot be guaranteed). Similar to penalty-based approach, additional criteria can be used to rank/filter alternative paths (e.g., using a ``goodness cost function''~\cite{jones2012method}). The computational cost to compute alternative paths using plateaus consists of generating two shortest path trees (e.g., using Dijkstra's algorithm) and joining the two trees. The latter can be done in time linear to the size of the tree~\cite{jones2012method}. Thus, the total cost is dominated by the two Dijkstra searches.

\subsection{Dissimilarity}\label{sec:dissimilarity}

Some existing works~\cite{chondrogiannis2020finding,chondrogiannis2018finding,liu2017finding} specifically define a dissimilarity function $dis(p,P)$ to compute the dissimilarity of a candidate path $p$ to a set of  paths $P$. The aim is to iteratively add paths to the result set $P$ in ascending order of their lengths as long as they are sufficiently dissimilar to the previously selected paths $P$. Specifically,  a path $p$ is added to $P$ only if $dis(p,P)>\theta$ where $\theta$ is a user-defined dissimilarity threshold. As a result, the $k$ paths reported to the user are significantly different from each other and are short. 

\TKDEOnly{
\begin{figure*}[bpth!]
	\centering
    \subfigure[For a user selected source and target, alternative routes by each approach are displayed]{\includegraphics[width=0.6\textwidth]{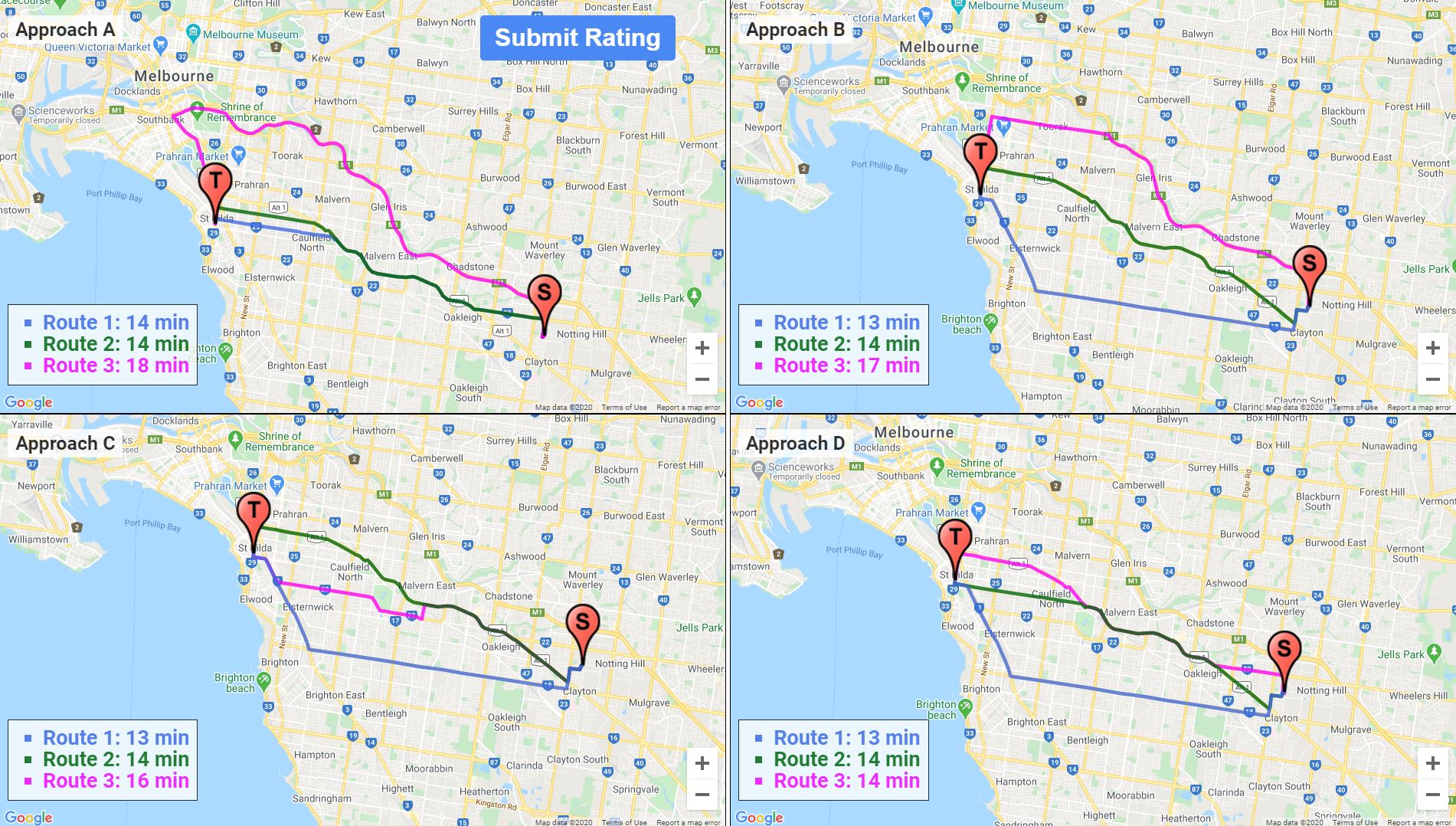}
    \label{fig:demo:b}}
      \subfigure[Feedback form for Melbourne]{\includegraphics[width=0.285\textwidth]{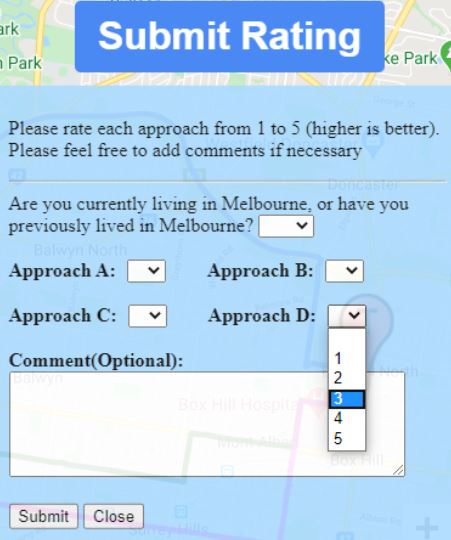}
    \label{fig:demo:c}}
  \caption{User Interface}
 	\label{fig:demo}
\end{figure*}
}

\arxivOnly{

\begin{figure*}[bpth!]
	\centering
  \subfigure[A user selects source and target locations]{\includegraphics[width=0.48\textwidth]{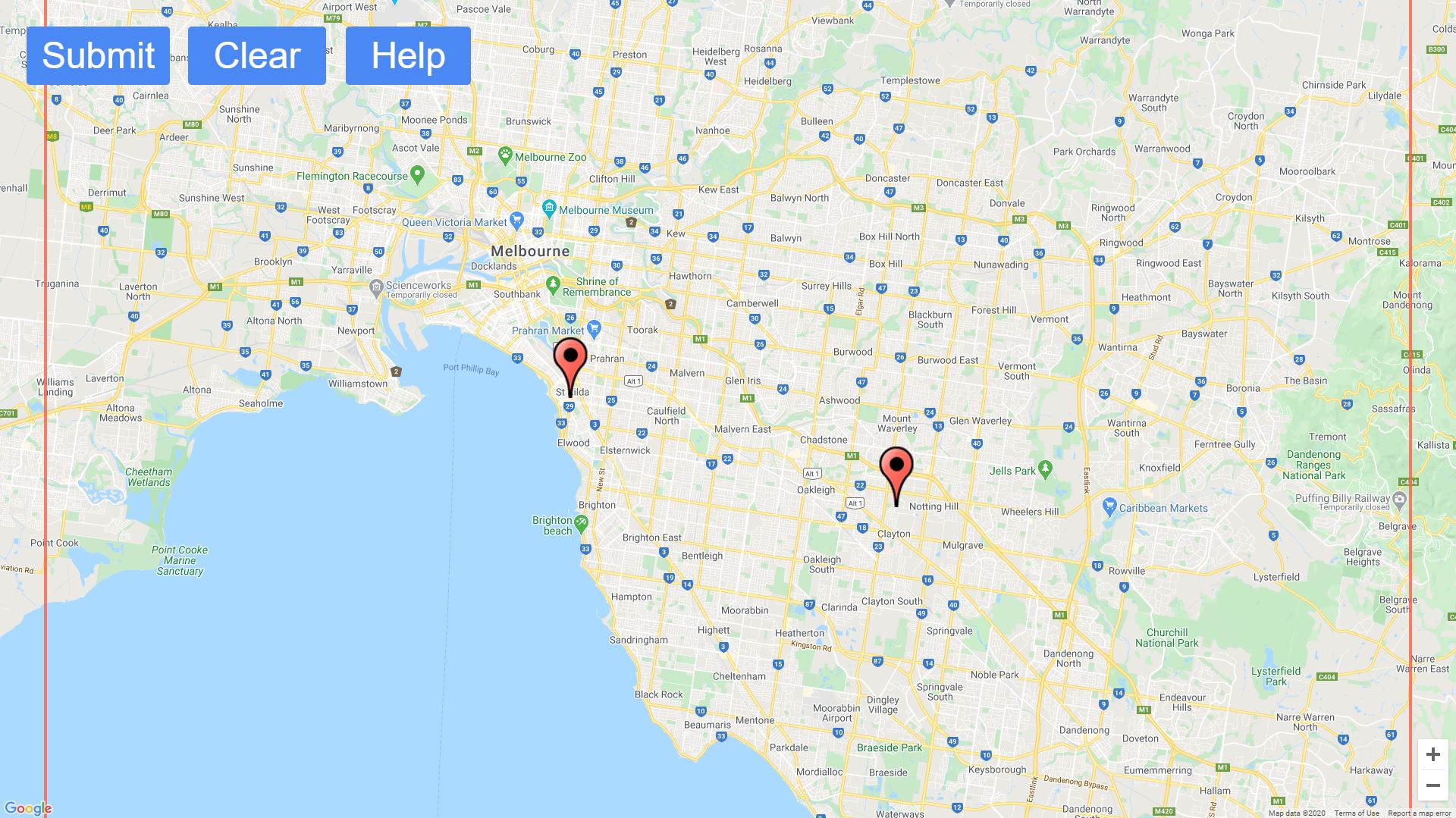}
  \label{fig:demo:a} }
    \subfigure[Alternative routes by each approach are displayed]{\includegraphics[width=0.48\textwidth]{figures/query_page.JPG}
    \label{fig:demo:b}}
  \caption{User Interface}
 	\label{fig:demo}
\end{figure*}

\begin{figure}[bpth!]
\centering
\includegraphics[width=0.3\textwidth]{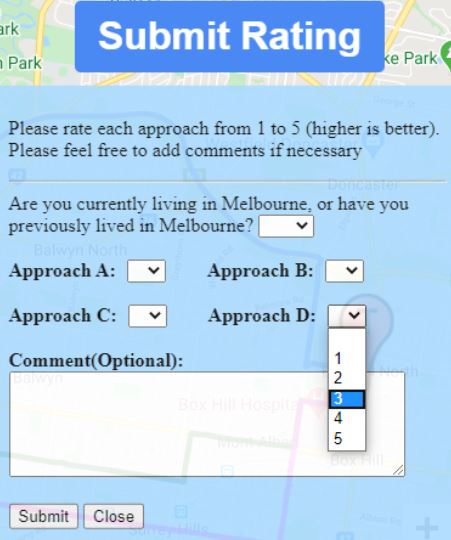}
  \caption{Feedback form for Melbourne}
    \label{fig:demo:c}
\end{figure}
}
The advantage of this approach is that it guarantees that the alternative paths are sufficiently dissimilar to each other (as defined by the parameter $\theta$). However, this approach does not guarantee that the generated alternative paths are free of small unnecessary detours. This can be addressed by having additional filtering criteria to prune the paths that do not meet certain criteria. A major disadvantage of this approach is that the problem is NP-hard~\cite{chondrogiannis2020finding, liu2017finding}. The existing studies have proposed several approximate algorithms. However, many of these techniques still appear to be quite slow taking tens of seconds to report alternative paths on a city-scale road network. 

In this study, we use SSVP-D+~\cite{chondrogiannis2018finding} which has been shown to generate good quality alternative routes and has reasonable computation cost. The basic idea is to use \emph{via-nodes} to generate paths. A path (called via-path) generated using a via-node $u$ is the concatenation of $sp(s,u)$ and $sp(u,t)$ where $sp(x,y)$ denotes the shortest path from $x$ to $y$. To efficiently compute $sp(s,u)$ and $sp(u,t)$, similar to the plateaus-based approach, two shortest path trees are constructed rooted at $s$ and $t$, respectively. The algorithm iteratively selects via-nodes in an ascending order of their via-paths lengths. A via-path is added to the result set only if its dissimilarity to the existing paths in $P$ is greater than the threshold $\theta$.

\subsection{Other techniques}

Yen's algorithm~\cite{yen1971finding} can be used to compute $k$ shortest paths from $s$ to $t$. However, these $k$ shortest paths are all expected to be very similar to each other. Thus, Yen's algorithm is not suitable for generating alternative paths if applied trivially. However, some existing techniques (e.g., see~\cite{chondrogiannis2015alternative}) use Yen's algorithm to incrementally generate shortest paths and apply filtering techniques to prune the paths that do not meet certain criteria. 
Pareto optimal \cite{barth2019alternative, barth2019alternativeB} paths (i.e., skyline paths) report the paths that are not \emph{dominated} by any other path according to given criteria (e.g., distance, travel time and ease of traveling). Some techniques~\cite{abraham2013alternative,luxen2015candidate} use \emph{via-nodes} to generate alternative paths (e.g., the SSVP-D+ techniques discussed in Section~\ref{sec:dissimilarity}). Such techniques identify interesting via-nodes in the road network and then apply different filtering/ranking criteria to generate the top-$k$ alternative paths. 
Some techniques compute alternative graphs~\cite{DBLP:journals/arxi,bader2011alternative} that concisely represent a union of paths from source to target. The alternative routes then can be computed using different ranking functions depending on user preferences.
There also exists techniques~\cite{skoumas2015knowledge,josse2017knowledge,josse2015tourismo} that study an orthogonal problem of using a road network enriched with additional information to report routes that consider not only the distance but also other criteria such as scenic value, tourist attractiveness and popularity etc.

\section{Demonstration System}\label{sec:system}
We use the source code from our previous visualisation~\cite{li2018visualising} and significantly extend it to create a web-based demonstration system\footnote{\url{http://aamircheema.com/routing/demo.html}}. The system
consists of the following major components: 1) road network constructor; 2) user-interface; 3) query processor. We briefly describe each component below.

\noindent \underline{\textbf{Road Network Constructor:}}
The road network constructor takes a rectangular area as input and extracts the road network data from OpenStreetMap (OSM) that lies within the input rectangle. 
First,  we export the raw OSM data using Geofabrik\footnote{\url{http://download.geofabrik.de/}}. 
Then, we filter the data that lies in the input rectangle. Finally, we parse the content of this raw OSM data to generate the road network data to be used by the approaches. Specifically, we extract tuples where each tuple represents an edge of the road network along with its end vertices and edge weight (travel time). The travel time is obtained by dividing the length of the edge with the maximum speed along the edge. In real-world scenarios, vehicles may need to stop at intersections, wait at traffic lights or slow down while turning even when there is no congestion on the roads. Thus, estimating the travel time using the maximum speed is not realistic. To better simulate real-world scenarios, for each road segment that is not a freeway/motorway, we multiply the edge weight (travel time) by 1.3. Our trials showed that this results in a reasonably good estimate of actual travel time when the roads have no congestion (e.g., compared with the travel time estimated by Google Maps at 2:00 am).

\TKDEOnly{
\renewcommand{\arraystretch}{1.2}       
\begin{table*}[!hbpt]
\centering
\begin{tabular}{|c|c|l|c|c|c|c|c|} \hline
	    & &  & \textbf{Google Maps}& \textbf{Plateaus}& \textbf{Dissimilarity}& \textbf{Penalty}& \textbf{\#Responses} \\ \hline

     \multirow{12}{*}{\rotatebox[origin=c]{90}{\textbf{All Cities}}} & \multirow{4}{*}{\rotatebox[origin=c]{90}{\textbf{All}}} & \textbf{All responses} & 3.37 (1.33)	& 3.44 (1.28)	& 3.43 (1.28)	& 	\textbf{3.53} (1.32)	& 520 \\ \cline{3-8}
	& & \textbf{Small Routes (0, 10] (mins)}   & 3.54 (1.28)&	3.49 (1.30)&	3.52 (1.30)&	\textbf{3.69} (1.30)&	143 \\ \cline{3-8}
	&  & \textbf{Medium Routes (10, 25] (mins)}  & 3.31 (1.37)&	3.34 (1.26)& \textbf{3.43} (1.25)&	3.42 (1.35)& 246
 \\ \cline{3-8}
     & & \textbf{Long Routes (25, 80] (mins)}  & 3.29 (1.29)&	\textbf{3.58} (1.29)&	3.33 (1.33)&	3.56 (1.26)& 131
 \\
	\cline{2-8}
	& \multirow{4}{*}{\rotatebox[origin=c]{90}{\textbf{Residents}}}  &	\textbf{All residents} & \textbf{3.55} (1.32)&	3.43 (1.26)&	3.44 (1.30)&	3.49 (1.29)&	334 \\
	\cline{3-8}
    & &	\textbf{Small Routes (0, 10] (mins)}   & 3.54 (1.29)&	3.38 (1.31)&	3.47 (1.37)&	\textbf{3.61} (1.36)& 110
 \\ \cline{3-8}
    & &	\textbf{Medium Routes (10, 25] (mins)}  & \textbf{3.54} (1.35)&	3.43 (1.25)&	3.43 (1.26)&	3.37 (1.30)&167
 \\ \cline{3-8}
    & &  \textbf{Long Routes (25, 80] (mins)}   & \textbf{3.60} (1.32)&	3.51 (1.21)&	3.40 (1.29)&	3.58 (1.13)& 57 \\
	\cline{2-8}
	& \multirow{4}{*}{\rotatebox[origin=c]{90}{\textbf{Non-resd.}}} & \textbf{All Non-residents} & 3.04 (1.28)&	3.47 (1.28)&	3.41 (1.25)&	\textbf{3.60} (1.36)&186 \\ \cline{3-8}
	& & \textbf{Small Routes (0, 10] (mins)}   & 3.55 (1.28)&	3.85 (1.20)&	3.70 (1.02)&	\textbf{3.94} (1.06)& 33
 \\ \cline{3-8}
	& & \textbf{Medium Routes (10, 25] (mins)}  & 2.82 (1.29)&	3.15 (1.26)&	3.43 (1.22)&	\textbf{3.53} (1.46)& 79
 \\ \cline{3-8}
    & & \textbf{Long Routes (25, 80] (mins)}   & 3.05 (1.23)&	\textbf{3.64} (1.36)&	3.27 (1.37)&	3.53 (1.36)& 74
 \\	\hline

    \multicolumn{2}{|c|}{ \multirow{3}{*}{\textbf{Melbourne}}} & \textbf{All responses} & 3.38 (1.33) & \textbf{3.63} (1.25) & 3.58 (1.29) &	3.57 (1.17) & 237 \\ \cline{3-8}
	\multicolumn{2}{|l|}{} &  \textbf{Residents only} & 3.55 (1.28) & 3.69(1.17)	& \textbf{3.71} (1.23) &	3.67 (1.12) & 156 \\
	\cline{3-8}
	\multicolumn{2}{|l|}{} &   \textbf{Non-residents only} & 3.05 (1.37) &	\textbf{3.51} (1.38) &	3.35 (1.37) &	3.37 (1.25)	 & 81 
 \\	\hline

      \multicolumn{2}{|c|}{ \multirow{3}{*}{\textbf{Dhaka}}} & \textbf{All responses} & 3.36 (1.29) & 3.45 (1.27) &	3.35 (1.26)	& \textbf{3.47} (1.42) &	155 \\ \cline{3-8}
	\multicolumn{2}{|l|}{} &	\textbf{Residents only} & \textbf{3.52} (1.29) & 3.36 (1.28) &	3.30 (1.33)	& 3.43 (1.40) &	112 \\
	\cline{3-8}
	\multicolumn{2}{|l|}{} & \textbf{Non-residents only} & 2.95 (1.21) &	\textbf{3.67} (1.25)	& 3.47 (1.03) &	3.58 (1.48) & 43
 \\ \hline

     \multicolumn{2}{|c|}{ \multirow{3}{*}{\textbf{Copenhagen}}}  & \textbf{All responses} & 3.34 (1.37) &	3.09 (1.28) &	3.24 (1.28) &	\textbf{3.53} (1.44) &	128 \\ \cline{3-8}
	\multicolumn{2}{|l|}{} &	\textbf{Residents only} & \textbf{3.56} (1.47) & 2.92 (1.28) &	3.03 (1.29)	& 3.17 (1.43) &	66 \\
	\cline{3-8}
	\multicolumn{2}{|l|}{}  & \textbf{Non-residents only} & 3.10 (1.22) &	3.27 (1.27) & 3.47 (1.24) &	\textbf{3.92} (1.36)	& 62
 \\	\hline
\end{tabular}
	\caption{Average rating (AVG) and standard deviation of observations $sd$ for each approach shown as AVG ($sd$).}
		\label{table:rd}
\end{table*}

}

\noindent \underline{\textbf{User Interface:}}
The user interface is a dynamic web page created using HTML, Javascript, and JQuery. It has two main functionalities: (1) sending the user's query request to the back-end, and (2) interacting with Google Maps API\footnote{\url{https://developers.google.com/maps/documentation/}} 
to plot the routes on Google Maps. A user can click anywhere on the map within a specified rectangular area to pick two markers corresponding to the source $s$ and target $t$, 
\TKDEOnly{respectively.}
\arxivOnly{respectively (see Fig.~\ref{fig:demo:a}).} 
When the user presses the ``Submit'' button, $s$ and $t$ are sent to the back-end which computes the alternative routes generated by the Penalty, Plateaus and Dissimilarity techniques. The routes generated by Google Maps are also obtained by calling its API. The routes generated by each of the four approaches are displayed in a new window (see Fig.~\ref{fig:demo:b}).
The approaches are named A-D (A: Google Maps, B: Plateaus, C: Dissimilarity and D: Penalty). This is to hide the identities of the approaches from the users, to avoid any biases or preconceived notions.

The user can rate each approach by clicking on the ``Submit Rating'' button which opens a form as shown in 
Fig.~\ref{fig:demo:c}. The form requires the user to rate each approach from 1-5 (higher is better). We also ask the users whether they are currently living (or have lived) in the respective city. This enables us to analyse the results based on the ratings received by residents vs non-residents for each city. 

\arxivOnly{
\renewcommand{\arraystretch}{1.2}       
\begin{table*}[!hbpt]
\centering
\begin{tabular}{|c|c|l|c|c|c|c|c|} \hline
	    & &  & \textbf{Google Maps}& \textbf{Plateaus}& \textbf{Dissimilarity}& \textbf{Penalty}& \textbf{\#Responses} \\ \hline

     \multirow{12}{*}{\rotatebox[origin=c]{90}{\textbf{All Cities}}} & \multirow{4}{*}{\rotatebox[origin=c]{90}{\textbf{All}}} & \textbf{All responses} & 3.37 (1.33)	& 3.44 (1.28)	& 3.43 (1.28)	& 	\textbf{3.53} (1.32)	& 520 \\ \cline{3-8}
	& & \textbf{Small Routes (0, 10] (mins)}   & 3.54 (1.28)&	3.49 (1.30)&	3.52 (1.30)&	\textbf{3.69} (1.30)&	143 \\ \cline{3-8}
	&  & \textbf{Medium Routes (10, 25] (mins)}  & 3.31 (1.37)&	3.34 (1.26)& \textbf{3.43} (1.25)&	3.42 (1.35)& 246
 \\ \cline{3-8}
     & & \textbf{Long Routes (25, 80] (mins)}  & 3.29 (1.29)&	\textbf{3.58} (1.29)&	3.33 (1.33)&	3.56 (1.26)& 131
 \\
	\cline{2-8}
	& \multirow{4}{*}{\rotatebox[origin=c]{90}{\textbf{Residents}}}  &	\textbf{All residents} & \textbf{3.55} (1.32)&	3.43 (1.26)&	3.44 (1.30)&	3.49 (1.29)&	334 \\
	\cline{3-8}
    & &	\textbf{Small Routes (0, 10] (mins)}   & 3.54 (1.29)&	3.38 (1.31)&	3.47 (1.37)&	\textbf{3.61} (1.36)& 110
 \\ \cline{3-8}
    & &	\textbf{Medium Routes (10, 25] (mins)}  & \textbf{3.54} (1.35)&	3.43 (1.25)&	3.43 (1.26)&	3.37 (1.30)&167
 \\ \cline{3-8}
    & &  \textbf{Long Routes (25, 80] (mins)}   & \textbf{3.60} (1.32)&	3.51 (1.21)&	3.40 (1.29)&	3.58 (1.13)& 57 \\
	\cline{2-8}
	& \multirow{4}{*}{\rotatebox[origin=c]{90}{\textbf{Non-resd.}}} & \textbf{All Non-residents} & 3.04 (1.28)&	3.47 (1.28)&	3.41 (1.25)&	\textbf{3.60} (1.36)&186 \\ \cline{3-8}
	& & \textbf{Small Routes (0, 10] (mins)}   & 3.55 (1.28)&	3.85 (1.20)&	3.70 (1.02)&	\textbf{3.94} (1.06)& 33
 \\ \cline{3-8}
	& & \textbf{Medium Routes (10, 25] (mins)}  & 2.82 (1.29)&	3.15 (1.26)&	3.43 (1.22)&	\textbf{3.53} (1.46)& 79
 \\ \cline{3-8}
    & & \textbf{Long Routes (25, 80] (mins)}   & 3.05 (1.23)&	\textbf{3.64} (1.36)&	3.27 (1.37)&	3.53 (1.36)& 74
 \\	\hline

     \multirow{12}{*}{\rotatebox[origin=c]{90}{\textbf{Melbourne}}} & \multirow{4}{*}{\rotatebox[origin=c]{90}{\textbf{All}}} & \textbf{All responses} & 3.38 (1.33) & \textbf{3.63} (1.25) & 3.58 (1.29) &	3.57 (1.17) & 237 \\ \cline{3-8}
	& & \textbf{Small Routes (0, 10] (mins)}   & 3.52 (1.18) & 3.56 (1.27) & 3.68 (1.16) & \textbf{3.87} (1.04) & 63 \\ \cline{3-8}
	&  & \textbf{Medium Routes (10, 25] (mins)}  & 3.43 (1.38) &	3.49 (1.25)	& \textbf{3.58} (1.27) &	3.39 (1.24) & 110
 \\ \cline{3-8}
     & & \textbf{Long Routes (25, 80] (mins)}   & 3.16 (1.37) & \textbf{3.94} (1.18) & 3.48 (1.44) & 3.56 (1.13)	& 64
 \\
	\cline{2-8}
	& \multirow{4}{*}{\rotatebox[origin=c]{90}{\textbf{Residents}}}  &	\textbf{All Melbourne residents} & 3.55 (1.28) & 3.69(1.17)	& \textbf{3.71} (1.23) &	3.67 (1.12) & 156 \\
	\cline{3-8}
    & &	\textbf{Small Routes (0, 10] (mins)}   & 3.51 (1.17) &	3.43 (1.28)	& 3.65 (1.25) &	\textbf{3.95} (1.00)	& 37
 \\ \cline{3-8}
    & &	\textbf{Medium Routes (10, 25] (mins)}  & 3.61 (1.28) &	3.72 (1.11) & \textbf{3.78} (1.13) &	3.56 (1.19) & 82
 \\ \cline{3-8}
    & &  \textbf{Long Routes (25, 80] (mins)}   & 3.46 (1.41) &	\textbf{3.89} (1.17)	& 3.59 (1.42) &	3.62 (1.06) &	37 \\
	\cline{2-8}
	& \multirow{4}{*}{\rotatebox[origin=c]{90}{\textbf{Non-resd.}}} & \textbf{All Non-residents} & 3.05 (1.37) &	\textbf{3.51} (1.38) &	3.35 (1.37) &	3.37 (1.25)	 & 81 \\ \cline{3-8}
	& & \textbf{Small Routes (0, 10] (mins)}   & 3.54 (1.21) &	3.73 (1.25) &	3.73 (1.04) &	\textbf{3.77} (1.11)	 & 26
 \\ \cline{3-8}
	& & \textbf{Medium Routes (10, 25] (mins)}  & 2.89 (1.55) &	2.82 (1.42)	& \textbf{3.00} (1.49) &	2.89 (1.29)	& 28
 \\ \cline{3-8}
    & & \textbf{Long Routes (25, 80] (mins)}   & 2.74 (1.23) &	\textbf{4.00} (1.21) & 3.33(1.47) &	3.48 (1.22) & 27
 \\	\hline
      \multirow{12}{*}{\rotatebox[origin=c]{90}{\textbf{Dhaka}}} & \multirow{4}{*}{\rotatebox[origin=c]{90}{\textbf{All}}} & \textbf{All responses} & 3.36 (1.29) & 3.45 (1.27) &	3.35 (1.26)	& \textbf{3.47} (1.42) &	155 \\ \cline{3-8}
	& & \textbf{Small Routes (0, 10] (mins)}   & 3.43 (1.33) & 3.55 (1.29)	& 3.43 (1.31) &	\textbf{3.62} (1.39) &	58 \\ \cline{3-8}
	&  & \textbf{Medium Routes (10, 20] (mins)}  & \textbf{3.33} (1.38) & 3.32 (1.31) &	3.27 (1.26)	& 3.30 (1.47) &	63
 \\ \cline{3-8}
     & & \textbf{Long Routes (20, 80] (mins)}  & 3.29 (1.09) &	3.50 (1.21) &	3.35 (1.18)	& \textbf{3.53} (1.40) &	34
 \\
	\cline{2-8}
	& \multirow{4}{*}{\rotatebox[origin=c]{90}{\textbf{Residents}}}  &	\textbf{All Dhaka residents} & \textbf{3.52} (1.29) & 3.36 (1.28) &	3.30 (1.33)	& 3.43 (1.40) &	112 \\
	\cline{3-8}
    & &	\textbf{Small Routes (0, 10] (mins)}   & 3.47 (1.31) & 3.49 (1.30) & 3.43 (1.35) &	\textbf{3.53} (1.41)	& 53
 \\ \cline{3-8}
    & &	\textbf{Medium Routes (10, 20] (mins)}  & \textbf{3.52} (1.35) & 3.31 (1.34)	& 3.19 (1.36) &	3.29 (1.43)	& 48
 \\ \cline{3-8}
    & &  \textbf{Long Routes (20, 80] (mins)}   & \textbf{3.73} (1.01) & 2.91 (0.83)	& 3.18 (1.17) &	3.55 (1.29) &	11 \\
	\cline{2-8}
	& \multirow{4}{*}{\rotatebox[origin=c]{90}{\textbf{Non-resd.}}} & \textbf{All Non-residents} & 2.95 (1.21) &	\textbf{3.67} (1.25)	& 3.47 (1.03) &	3.58 (1.48) & 43 \\ \cline{3-8}
	& & \textbf{Small Routes (0, 10] (mins)}   & 3.00 (1.58) & 4.20 (1.10) &	3.40 (0.89) & \textbf{4.60} (0.55) & 5
 \\ \cline{3-8}
	& & \textbf{Medium Routes (10, 20] (mins)}  & 2.73 (1.33) &	3.33 (1.23) & \textbf{3.53} (0.83) &	3.33 (1.63) & 15
 \\ \cline{3-8}
    & & \textbf{Long Routes (20, 80] (mins)}   & 3.09 (1.08) &	\textbf{3.78} (1.28) & 3.43 (1.20) &	3.52 (1.47) & 23
 \\\hline
     \multirow{12}{*}{\rotatebox[origin=c]{90}{\textbf{Copenhagen}}} & \multirow{4}{*}{\rotatebox[origin=c]{90}{\textbf{All}}} & \textbf{All responses} & 3.34 (1.37) &	3.09 (1.28) &	3.24 (1.28) &	\textbf{3.53} (1.44) &	128 \\ \cline{3-8}
	& & \textbf{Small Routes (0, 10] (mins)}   & \textbf{3.73} (1.49) &	3.14 (1.42) &	3.32 (1.62)	& 3.32 (1.67) &	22 \\ \cline{3-8}
	&  & \textbf{Medium Routes (10, 25] (mins)}  & 3.12 (1.33) &	3.14 (1.21)	& 3.33 (1.18) &	\textbf{3.58} (1.40)	& 73
 \\ \cline{3-8}
     & & \textbf{Long Routes (25, 80] (mins)}  & 3.55 (1.33) &	2.97 (1.38) & 3.00 (1.25) &	\textbf{3.58} (1.39)	& 33
 \\
	\cline{2-8}
	& \multirow{4}{*}{\rotatebox[origin=c]{90}{\textbf{Residents}}}  &	\textbf{All Copenhagen residents} & \textbf{3.56} (1.47) & 2.92 (1.28) &	3.03 (1.29)	& 3.17 (1.43) &	66 \\
	\cline{3-8}
    & &	\textbf{Small Routes (0, 10] (mins)}   & \textbf{3.60} (1.50) & 3.00 (1.41) & 3.25 (1.65) & 3.20 (1.70) & 20
 \\ \cline{3-8}
    & &	\textbf{Medium Routes (10, 25] (mins)}  & \textbf{3.43} (1.50) & 2.95 (1.27) & 2.95 (1.20) &	3.05 (1.31) & 37
 \\ \cline{3-8}
    & &  \textbf{Long Routes (25, 80] (mins)}   & \textbf{4.00} (1.32) & 2.67 (1.12) & 2.89 (0.60) &	3.56 (1.33)	& 9 \\
	\cline{2-8}
	& \multirow{4}{*}{\rotatebox[origin=c]{90}{\textbf{Non-resd.}}} & \textbf{All Non-residents} & 3.10 (1.22) &	3.27 (1.27) & 3.47 (1.24) &	\textbf{3.92} (1.36)	& 62 \\ \cline{3-8}
	& & \textbf{Small Routes (0, 10] (mins)}   & \textbf{5.00} (0.00) & 4.50 (0.71) &	4.00 (1.41) & 4.50 (0.71) &	2
 \\ \cline{3-8}
	& & \textbf{Medium Routes (10, 25] (mins)}  & 2.81 (1.06) &	3.33 (1.12) & 3.72 (1.03) &	\textbf{4.11} (1.30) & 36
 \\ \cline{3-8}
    & & \textbf{Long Routes (25, 80] (mins)}   & 3.38 (1.31) &	3.08 (1.47)	& 3.04 (1.43) &	\textbf{3.58} (1.44)	& 24
 \\	\hline
\end{tabular}
	\caption{Average rating (AVG) and standard deviation $sd$ for each approach shown as AVG ($sd$).}
		\label{table:rd}
\end{table*}
}
\arxivOnly{

\renewcommand{\arraystretch}{1.2}       
\begin{table*}[!hbpt]
\centering
\begin{tabular}{|c|l|c|c|c|c|c|c|c|c|c|} \hline
	    & & \multicolumn{2}{c}{\textbf{Google Maps}}& \multicolumn{2}{|c|}{\textbf{Plateaus}}& \multicolumn{2}{c}{\textbf{Dissimilarity}}& \multicolumn{2}{|c|}{\textbf{Penalty}} & \textbf{\#Responses} \\ \hline
	    & & AVG & MAX & AVG & MAX & AVG & MAX & AVG & MAX & \\ \cline{2-11}
	   
     \multirow{4}{*}{\rotatebox[origin=c]{90}{\textbf{All Cities}}} & \textbf{All responses} & 0.353 (0.15)& 0.800 &	0.382 (0.15)& 0.965&	\textbf{0.244} (0.06)& \textbf{0.331}&	0.395 (0.22)& 0.995 &	520 \\ 
     \cline{2-11}
	& \textbf{Small Routes}   & 0.325 (0.21)& 0.800&	0.434 (0.19)& 0.863&	\textbf{0.236} (0.12)& \textbf{0.322}&	0.479 (0.27)& 0.980& 143 \\ 
	\cline{2-11}
	& \textbf{Medium Routes}  & 0.349 (0.15)& 0.675&	0.356 (0.13)& 0.965&	\textbf{0.242} (0.10)& \textbf{0.328}&	0.381 (0.21)& 0.988& 246\\ 
	\cline{2-11}
     & \textbf{Long Routes}  & 0.382 (0.15)& 0.703&	0.377 (0.13)& 0.728&	\textbf{0.254} (0.10)& \textbf{0.331}&	0.337 (0.2)& 0.995& 131 \\
    \hline
	    
     \multirow{4}{*}{\rotatebox[origin=c]{90}{\textbf{Melbourne}}} & \textbf{All responses} & 0.335 (0.14) & 0.716 & 0.380 (0.16) & 0.965 & \textbf{0.247} (0.05) & \textbf{0.328} & 0.376 (0.26) & 0.995  & 237 \\ \cline{2-11}
	& \textbf{Small Routes}   & 0.310 (0.20) & 0.716 & 0.457 (0.20) & 0.862 & \textbf{0.240} (0.12) & \textbf{0.302} & 0.526 (0.30) & 0.980 & 63 \\ \cline{2-11}
	&  \textbf{Medium Routes}  & 0.336 (0.14) & 0.658 & 0.348 (0.14) & 0.965 & \textbf{0.242} (0.09) & \textbf{0.328} & 0.352 (0.26) & 0.988 & 110 \\ \cline{2-11}
     & \textbf{Long Routes}   & 0.356 (0.14) & 0.584 & 0.362 (0.13) & 0.695 & \textbf{0.261} (0.09) & \textbf{0.325} & 0.280 (0.19) & 0.995 & 64 
 \\	\hline
      \multirow{4}{*}{\rotatebox[origin=c]{90}{\textbf{Dhaka}}} &  \textbf{All responses} & 0.375 (0.15) & 0.703 & 0.408 (0.14)& 0.863 & \textbf{0.234} (0.07)& \textbf{0.331} & 0.391 (0.22)& 0.963 & 155 \\ \cline{2-11}
	& \textbf{Small Routes}   & 0.358 (0.18)& 0.697 & 0.404 (0.18)& 0.863 & \textbf{0.233} (0.12)& \textbf{0.322} & 0.406 (0.25)& 0.963 &	58 \\ \cline{2-11}
	& \textbf{Medium Routes}  & 0.371 (0.16)& 0.675 & 0.407 (0.12)& 0.752 & \textbf{0.234} (0.11)& \textbf{0.316} & 0.399 (0.19)& 0.875 & 63
 \\ \cline{2-11}
     & \textbf{Long Routes}  & 0.418 (0.22)& 0.703 & 0.418 (0.15)& 0.728 & \textbf{0.237} (0.11)& \textbf{0.331} & 0.353 (0.24)&	0.905 & 34
 \\
 \hline
     \multirow{4}{*}{\rotatebox[origin=c]{90}{\textbf{Copenhagen}}} & \textbf{All responses} & 0.365 (0.14)& 0.800 & 0.357 (0.14)& 0.845 & \textbf{0.249} (0.06)& \textbf{0.329} & 0.435 (0.12)& 0.845 &	128 \\ \cline{2-11}
	& \textbf{Small Routes}   & 0.348 (0.24) & 0.800 & 0.465 (0.22)& 0.845 & \textbf{0.234} (0.13)& \textbf{0.309} & 0.533 (0.17)& 0.845 & 22 \\ 
	\cline{2-11}
	& \textbf{Medium Routes}  & 0.351 (0.14)& 0.587 & 0.323 (0.12)& 0.650 & \textbf{0.248} (0.08)& \textbf{0.324} & 0.408 (0.10)& 0.650 & 73\\ 
	\cline{2-11}
     & \textbf{Long Routes}  & 0.404 (0.11)& 0.556 &	0.364 (0.10)& 0.641 & \textbf{0.257} (0.10)& \textbf{0.329}	& 0.431 (0.9)&  0.641 & 33 \\
    \hline

\end{tabular}
	\caption{Average (AVG) and maximum (MAX) $Sim(T)$ for each approach. Standard deviation is shown in parenthesis next to the average.}
		\label{table:sim}
\end{table*}

}

\noindent \underline{\textbf{Query Processor}:}
  Input to the query processor is a pair of source and target locations each represented by longitude and latitude. First, the query processor performs geo-coordinate matching and selects the closest vertices from the OSM data to the source and target locations, respectively. 
      Then, the query processor  computes the alternative routes from the source location to the target location using the three techniques we implemented (Penalty, Plateaus and Dissimilarity). It also calls Google Maps API to get the alternative routes generated by Google Maps.
    For each of the routes generated by these four approaches, the query processor computes its travel time by using the OSM data. Each travel time is rounded to display time in minutes. Finally, the routes generated by each approach are passed to Google Maps API to display these routes using different colors so that they are easily distinguishable.

\noindent
\underline{\textbf{Parameter  Details:}} The three approaches that we implemented (Penalty, Plateaus and Dissimilarity) use some parameters. As suggested in~\cite{bader2011alternative}, for the Penalty approach, the penalty that we apply to each edge is $1.4$, i.e., the edge weight is multiplied by $1.4$. For the Plateaus and Dissimilarity approaches, we use the upper bound~\cite{abraham2013alternative} to be $1.4$ which ensures that the travel time of any alternative path reported by these approaches is not higher than $1.4$ times  the travel time of the fastest path. The dissimilarity threshold $\theta$ for the Dissimilarity approach is set to $0.5$ as suggested  in the relevant studies~\cite{chondrogiannis2018finding,chondrogiannis2020finding}. We tried several other values for each of the above mentioned parameters to confirm that the chosen values are appropriate.

\section{User Study}\label{sec:experiments}

We created a web page\footnote{\url{http://aamircheema.com/routing}}
explaining the background of this research and providing instructions to the participants mostly living in Melbourne, Dhaka and Copenhagen. The participants were able to provide feedback using any internet enabled device on the road networks of Melbourne, Dhaka or Copenhagen.  
For the sake of this study, we say a participant is a resident if he/she was living (or had lived) in the city for which he/she completed the survey. We also sent requests to non-residents to get their ratings for each approach based on their perceived route quality. The residents were requested to select the source and target locations for the routes familiar to them (although we cannot guarantee if they always did so). We made sure that none of the approaches tries to avoid toll roads and we told the participants to ignore toll charges on the roads.

\subsection{Evaluation on participants' ratings}

Table~\ref{table:rd} shows the results (mean rating and standard deviation of observations) for each approach for all $520$ responses we received for the three cities. We also show the results grouped by different cities as well as different groups of participants (residents vs non-residents). We also group the responses based on the lengths of the shortest routes from source to target locations. Specifically, small routes correspond to all the responses where the fastest travel time from $s$ to $t$ was at most $10$ minutes. Medium and long routes correspond to the responses with fastest travel time from $s$ to $t$ within ranges $(10,25]$ and $(25,80]$ minutes, respectively.
\TKDEOnly{ Due to the space limitations, we do not show the results grouped by route length for each city, separately. The interested readers are referred to the extended version~\cite{arxiv-demo} for these results.}
Standard error of mean is not shown but can be easily computed by using the standard deviation and the number of responses. The highest mean rating  for each group is shown in bold.

 The overall results considering all responses from all three cities show that Penalty achieves the highest mean rating although the other three approaches are not too far behind.  For long routes, Plateaus has the highest mean rating whereas for medium and small routes Dissimilarity and Penalty have the highest mean ratings, respectively. Considering only the responses from the residents, Google Maps has the highest mean rating.
 Interestingly, Google Maps consistently received lower mean ratings from non-residents as compared to the residents. This indicates that the routes provided by Google Maps may be perceived to be of poor quality by non-residents when in fact they may not be necessarily so.
  Considering all responses, Plateaus has the highest mean rating for Melbourne road network whereas Penalty  performs the best for Dhaka and Copenhagen road networks. On the other hand, considering only the residents, Dissimilarity achieves the highest mean rating for Melbourne whereas Google Maps received the highest rating  for Dhaka and Copenhagen. However, as discussed next, these results are not statistically significant.
  
  \TKDEOnly{

\renewcommand{\arraystretch}{1.2}       
\begin{table*}[!hbpt]
\centering
\begin{tabular}{|c|l|c|c|c|c|c|c|c|c|c|} \hline
	    & & \multicolumn{2}{c}{\textbf{Google Maps}}& \multicolumn{2}{|c|}{\textbf{Plateaus}}& \multicolumn{2}{c}{\textbf{Dissimilarity}}& \multicolumn{2}{|c|}{\textbf{Penalty}} & \textbf{\#Responses} \\ \hline
	    & & AVG & MAX & AVG & MAX & AVG & MAX & AVG & MAX & \\ \cline{2-11}
	   
     \multirow{4}{*}{\rotatebox[origin=c]{90}{\textbf{All Cities}}} & \textbf{All responses} & 0.353 (0.15)& 0.800 &	0.382 (0.15)& 0.965&	\textbf{0.244} (0.06)& \textbf{0.331}&	0.395 (0.22)& 0.995 &	520 \\ 
     \cline{2-11}
	& \textbf{Small Routes}   & 0.325 (0.21)& 0.800&	0.434 (0.19)& 0.863&	\textbf{0.236} (0.12)& \textbf{0.322}&	0.479 (0.27)& 0.980& 143 \\ 
	\cline{2-11}
	& \textbf{Medium Routes}  & 0.349 (0.15)& 0.675&	0.356 (0.13)& 0.965&	\textbf{0.242} (0.10)& \textbf{0.328}&	0.381 (0.21)& 0.988& 246\\ 
	\cline{2-11}
     & \textbf{Long Routes}  & 0.382 (0.15)& 0.703&	0.377 (0.13)& 0.728&	\textbf{0.254} (0.10)& \textbf{0.331}&	0.337 (0.2)& 0.995& 131 \\
    \hline

\end{tabular}
	\caption{Average (AVG) and maximum (MAX) $Sim(T)$ for each approach. Standard deviation is shown in parenthesis next to the average.}
		\label{table:sim}
\end{table*}

}
 
 We conducted one-way repeated measures ANOVA tests for different categories of respondents. Given a null hypothesis of no statistically significant difference in 
mean ratings of the four approaches, the p-value is relatively high. e.g., results for Melbourne, Dhaka and Copenhagen considering all responses were $[F(3,944)=2.197, p=0.087]$, $[F(3,616)=0.502, p=0.68]$ and $[F(3,508)=2.58, p=0.054]$, respectively, and considering residents only $[F(3,620)=0.592, p=0.62]$, $[F(3,444)=0.843, p=0.471]$ and $[F(3,260)=2.56, p=0.057]$, respectively. The results suggest that, at $p<0.05$ level, there is no evidence that the null hypothesis is false, i.e., there is no credible evidence that the four approaches receive different ratings on average.

\subsection{Evaluation on route similarity}

We also compare the approaches on the similarity of the alternative routes reported by them. 
Similar to many existing works (e.g., see~\cite{liu2017finding} and references therein), we use Eq~\eqref{eq:sim} to compute  similarity of a set of reported routes $T$ where $|X \cap Y|$ (resp.  $|X\cup Y|$) denote the total length of the overlap (resp. union) of two routes $X$ and $Y$. 

\begin{equation}\label{eq:sim}
Sim(T) = \argmax_{\forall(X,Y)\in T\times T~s.t. X\neq Y} \frac{|X\cap Y|}{|X\cup Y|}
\end{equation}

Table~\ref{table:sim} shows the results for \TKDEOnly{all}\arxivOnly{different} cities for the queries for which each approach reports $3$ alternative routes (including the shortest path). \TKDEOnly{The results grouped by each individual city are shown in the extended version~\cite{arxiv-demo}.}
For each category, we report average and maximum of $Sim(T)$ for the queries belonging to the category. As expected, Dissimilarity reports the routes with the least similarity because it specifically prunes the routes that have high similarity. Penalty and Plateaus sometimes report routes with a very high $Sim(T)$ (e.g., maximum $Sim(T)$ for Penalty is 0.995 for long routes implying that, for at least one query, two routes reported by Penalty were very similar). Nevertheless, we remark that additional constraints can be easily integrated in these approaches to prune the routes that have similarity higher than a given threshold.

\subsection{Limitations of the study}

While we have tried our best to be as fair as possible to all the approaches, certain factors beyond our control may have affected the participants' ratings. Below we list some of these.

\noindent
\textbf{Different data used by Google Maps and other approaches:}
One of the major factors potentially affecting the participants' ratings is that the data used by Google Maps and the OpenStreetMap (OSM) data used by the other three approaches are different. To estimate the travel time, Google Maps uses real-time traffic data (or historical data for queries issued at future dates/times). To minimize the impact of real-time traffic or historical traffic data, we call Google Maps API to retrieve the routes at 2:00 am on the next day (assuming minimal traffic on roads at that time). However, the travel time estimation for the same route may still be different for Google Maps and OpenStreetMap data. We observe that this difference in the underlying data affects the routes generated by Google Maps and the other approaches.

\begin{figure*}[th!]
	\centering
  \subfigure[Routes by Google Maps]{\includegraphics[width=0.48\textwidth]{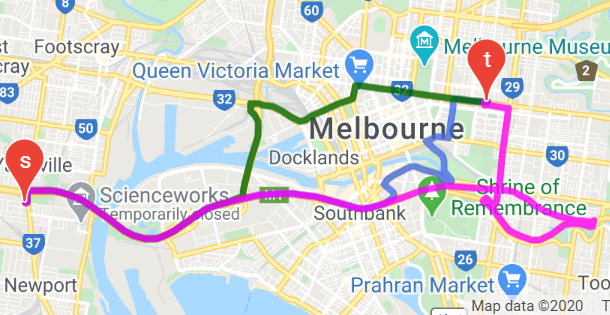}
  \label{fig:worst:Google} }
    \subfigure[Routes by Plateaus]{\includegraphics[width=0.48\textwidth]{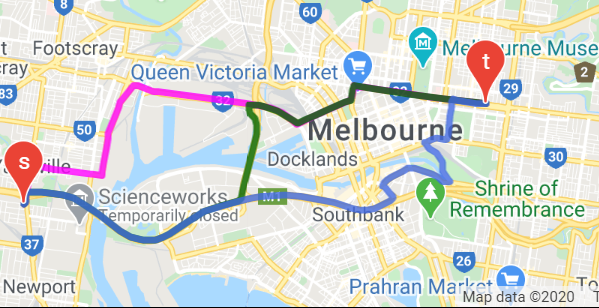}
    \label{fig:best:Google}}
  \caption{Blue and Green routes by both approaches are the same.  The pink route by Google Maps looks more complicated and has longer travel time than the pink route by Plateaus when OpenStreetMap data is used but has smaller travel time when Google Maps data is used to compute the travel time.}
 	\label{fig:Google}
\end{figure*}

Consider the example given in Fig.~\ref{fig:Google} that shows the alternative routes generated by Google Maps and Plateaus for the same pair of source $s$ and target $t$. While the blue and green routes returned by both approaches are the same, the pink route returned by Google Maps looks significantly more complicated and appears to have detours. We carefully looked into the pink routes provided by the two approaches. It turns out that if the data from OpenStreetMap is used, the travel time of the pink route by Google Maps is a few minutes higher than that of the pink route by Plateaus. However, when Google Maps data is used to obtain the travel times\footnote{We  manually inserted waypoints (i.e., intermediate destinations) to force Google Maps to generate the pink path given by Plateaus and obtained the travel time of this route as reported by Google Maps.}, the travel time of the pink route taken by Plateaus is a few minutes higher than that of the pink route by Google Maps. A user looking at the routes in Fig.~\ref{fig:Google} is likely to give a higher rating to Plateaus but this may be unfair because the two approaches are essentially using different underlying data.

\noindent
\textbf{Apparent detours that are not:} Due to the complex structure of road networks, a participant may incorrectly assume that a route provided by an approach has a detour. In the example of Fig.~\ref{fig:worst:Google}, the pink route by Google Maps appears to have a detour, e.g., one may assume that the route should have turned left at around ``Shrine of Remembrance'' instead of the detour. However, this is not a detour because the route goes through a tunnel and there is no left turn available near ``Shrine of Remembrance'', i.e., the path returned by Google Maps is a reasonable path. Unless a user is familiar with these roads and/or carefully looks at the road structure (e.g., by zooming in), he/she may perceive it as a detour and give a lower rating. We remark that this does not  only negatively affect the ratings received by Google Maps. Other approaches may also be negatively affected by such scenarios where a user may incorrectly assume a reasonable route to have detours.

\noindent
\textbf{Biases to favorite routes/navigation systems:}
A participant's rating may be biased by his/her favorite routes (or his/her favorite navigation system). E.g., a person who regularly uses Waze to travel from home to office may perceive the routes that are similar to those provided by Waze to be of better quality, although this may not be necessarily true. For example, one participant submitted the following comment: \textit{``no route using Blackburn rd''}. We believe that the route via Blackburn road is his/her favorite route which was not returned by any of the approaches. Consequently, the  maximum rating the user gave to any approach in this case was $3$.

\noindent
\textbf{Additional filtering/ranking criteria are not considered:}
For the approaches that we implemented (Plateaus, Dissimilarity and Penalty), we
could choose to use some additional filtering/ranking criteria to refine the top-$k$ alternative routes. For example, 
we could improve the routes generated by Penalty and Plateaus by pruning the alternative routes that have very high similarity to the other routes. Similarly, we could filter the routes in Penalty and Dissimilarity approaches that did not satisfy local optimality~\cite{abraham2013alternative}. Some comments from the participants also point out that, at least some, users consider certain factors to be important which we did not consider in our implementation but can be easily  included. 

For example, some comments from the participants were: \textit{``Approach C provides paths with less turns}'';
\textit{``less zig-zag is better''}; and \textit{``highest rated path follows wide roads''}. This indicates that these users perceive the routes to be better if they have fewer turns or are using wider roads (which probably means roads with more lanes). 
Since Google Maps is a widely used commercial product, we believe that they would have spent significant time and resources to identify such potentially important factors and implement additional filtering/ranking criteria to report alternative routes. However, most of these filtering criteria can also be easily included for the other approaches but are not considered in this study.

Considering the factors mentioned above (and the other unforeseen factors), we recommend the readers to make any conclusions with caution. However, we believe that our study shows evidence that the three approaches that we implemented (Plateaus, Dissimilarity and Penalty) are comparable to Google Maps. We also received some comments from the participants indicating that they find the routes to be of similar quality. For example a participant commented: \textit{``I don't see these approaches as very distinct from each other.''}. Another participant sent a personal message to the authors stating that he is finding it hard to rank the approaches since they all seem to be of similar quality.

\section{Conclusions}\label{sec:conclusion}

This is the first detailed user study that compares some of the most popular approaches to generate alternative routes, including Google Maps which is one of the most widely used navigation applications. We develop a web-based demo system to conduct the user study on the road networks of three demographically diverse cities namely Melbourne, Dhaka and Copenhagen. Our results show that the routes produced by all four approaches considered in this study are perceived to be of similar quality by the users. We also identify the limitations of this study that might have potentially affected the ratings submitted by the participants. Despite the limitations of the study, we believe that it is fair to conclude that 
three popular approaches published in literature (Plateaus, Dissimilarity and Penalty) are promising and
the quality of routes generated by these is comparable to the routes provided by Google Maps.




\ifCLASSOPTIONcaptionsoff
  \newpage
\fi


\bibliographystyle{abbrv}
\bibliography{alternative_path}

\end{document}